\documentclass[%
 reprint, amsmath, amssymb, aps,]{revtex4-1}
\usepackage{booktabs}
\usepackage{graphicx}
\usepackage{dcolumn}
\usepackage{bm}
\usepackage{color}
\usepackage{url}
\usepackage[normalem]{ulem}
\usepackage{mathtools}
\usepackage{array}
\newcolumntype{P}[1]{>{\centering\arraybackslash}p{#1}}
\newcolumntype{M}[1]{>{\centering\arraybackslash}m{#1}}
\usepackage[caption=false]{subfig}
\usepackage{dcolumn}
\usepackage{graphicx, epsfig}
\usepackage[dvipsnames]{xcolor}
\usepackage{mathrsfs}
\usepackage{yhmath}
\usepackage[caption=false]{subfig}
\usepackage[normalem]{ulem}
\usepackage{mathtools}
\usepackage{bigints}
\usepackage{float}
\usepackage[colorlinks = true, linkcolor = purple, urlcolor  = blue, citecolor = blue, anchorcolor = blue]{hyperref}
\usepackage{float}
\usepackage{multirow}

\newcommand{\RN}[1]{%
  \textup{\uppercase\expandafter{\romannumeral#1}}%
}

\begin{document}

\preprint{APS/123-QED}

\title{Constraining Cosmic-Ray Acceleration and Escape in Middle-Aged Supernova Remnants with GeV–TeV Gamma-Ray Observations}

\author{Siyu Chen$^{1,3}$}
\email[]{amocily@gmail.com}

\author{Bing Theodore Zhang$^{2,3}$}
\email[]{zhangbing@ihep.ac.cn}

\author{Yi Xing$^4$}

\author{Siming Liu$^1$}

\author{Xunxiu Zhou$^1$}

\affiliation{$^1$ School of Physical Science and Technology, Southwest Jiaotong University, Chengdu 610031, People's Republic of China}
\affiliation{$^2$ Key Laboratory of Particle Astrophysics and Experimental Physics Division and Computing Center, Institute of High Energy Physics, Chinese Academy of Sciences, 100049 Beijing, China}
\affiliation{$^3$ TIANFU Cosmic Ray Research Center, Chengdu, Sichuan, China}
\affiliation{$^4$Shanghai Astronomical Observatory, Chinese Academy of Sciences‌}

\date{\today}
             
\begin{abstract}
In this work, we perform a systematic, time-dependent study of the gamma-ray emission from four representative middle-aged SNRs (W51C, IC~443, W44, W28), incorporating both CRs within the remnant shells and escaped CRs interacting with surrounding molecular clouds. 
We compare our results with GeV--TeV gamma-ray observations from Fermi-LAT, H.E.S.S., MAGIC, and LHAASO, including a dedicated analysis of the Fermi-LAT data for regions A and B associated with W28.
We find that the observed spectra favor steeper CR injection spectra with indices of \(\alpha\sim4.2\)--\(4.3\), maximum proton energies of $\sim$ \(100\)--\(300\) TeV, diffusion coefficients below the Galactic average, and CR acceleration efficiencies from a few to tens of percent. 
In particular, the VHE emission detected by LHAASO from W51C is more naturally explained by escaped CRs interacting with a nearby molecular cloud. We also investigate the contribution of escaped CRs to the VHE emission from IC~443, W44, and W28.
We further demonstrate that escaped CRs can substantially enhance the TeV neutrino flux from middle-aged SNRs, improving their prospects as potential neutrino sources.
These results provide new constraints on CR acceleration and escape in middle-aged SNRs and highlight the important role of escaped CRs in shaping their high-energy gamma-ray and neutrino emission.
\end{abstract}

\pacs{Valid PACS appear here}
\maketitle


\section{\label{sec:intro}Introduction}
Supernova remnants (SNRs) are widely regarded as the primary candidate sources of Galactic cosmic rays (CRs)~\cite[e.g.,][]{Bykov:2025}. 
Diffusive shock acceleration (DSA)—the first-order Fermi mechanism—is thought to operate in SNRs \cite{1978MNRAS.182..147B,Drury_1983,1977DoSSR.234.1306K,1978ApJ...221L..29B} : particles gain energy by scattering back and forth across the shock front via magnetic turbulence. 
Multi-wavelength observations have significantly strengthened the SNR–CR connection.
During the past two decades, several dozen SNRs have been detected in the GeV and TeV energy bands \cite{giuliani2024supernovaremnantsgammarays}.
Yet, despite this progress, whether SNRs are the main contributors to Galactic CRs remains open—and their ability to reach PeV energies lacks conclusive evidence.

Very-high-energy (VHE) gamma-ray observations of SNRs have played a pivotal role in establishing these objects as efficient hadron accelerators in the Galaxy. 
Over the past two decades, Imaging Atmospheric Cherenkov Telescopes (IACTs), such as H.E.S.S., MAGIC, have detected VHE gamma-rays from several SNRs \cite{2011ICRC....7..111B,2006A&A...449..223A,2018A&A...612A...7H,2012A&A...541A..13A}. 
While the origin of this VHE emission remains debated for some sources, two landmark results have provided clear evidence for hadron acceleration. Fermi-LAT detected the characteristic pion-decay spectral feature (the ``pion bump'') in IC 443 and W44~\cite{FermiLAT:2009kcy, Ackermann_2013,2010ApJ...712..459A}, unambiguously confirming the hadronic origin of their gamma-ray emission.
More recently, the LHAASO has detected W51C up to \(\sim 200\) TeV \cite{2024SciBu..69.2833C}. For IC~443, LHAASO resolved the gamma-ray emission into two distinct components: a point-like source C0 and an extended source C1 \cite{2026PhRvL.136p1002C}. The spectrum of C0 extends beyond \(\sim 30\) TeV without an apparent cutoff, providing compelling evidence that the SNR shock can accelerate protons to sub-PeV energies. The extended component C1 may be associated with IC~443 itself, the older neighboring SNR G189.6+3.3, or a pulsar wind nebula, and can be explained by either hadronic or leptonic models~\cite{2026PhRvL.136p1002C}.
LHAASO has extended this picture to PeV energies by detecting gamma-ray emission up to hundreds of TeV from the middle-aged shell-type SNRs G150.3+4.5 and $\gamma$-Cygni \cite{cao2026ultrahighenergygammarayimprintspev}. These observations indicate a hadronic origin associated with molecular clouds, reinforcing the view that SNRs are promising PeVatron candidates and likely major sources of Galactic CRs.

An important complication is that the highest-energy particles are generally expected to escape from the SNR system earlier than lower-energy particles. 
Consequently, the gamma-ray emission produced within the SNR shell may not preserve a clear signature of PeV particle acceleration, particularly at relatively late evolutionary stages.
A more reliable approach is therefore to search for gamma-ray emission produced when these escaped CRs interact with molecular clouds located outside, or in the vicinity of, the SNR. 
Such clouds can act as natural targets that record the propagation and interaction of the escaped particles.
The detection of spatially associated, hard-spectrum gamma-ray emission from molecular clouds can thus provide an indirect but powerful probe of whether an SNR has operated as a PeVatron, even when the shell emission itself no longer exhibits an unambiguous PeV signature.

In this study, we investigate the origin of high-energy gamma-ray emission from several archetypal middle-aged SNRs using available GeV-to-PeV observations, with particular emphasis on measurements in the VHE band. 
We perform detailed time-dependent modeling of CR acceleration and escape in SNRs, which tracks the evolution of both confined and escaped CRs, as well as their subsequent interactions with ambient gas clouds.
By fitting the model to the observed gamma-ray spectra, we constrain the key physical parameters, including the maximum acceleration energy, spectral index, and acceleration efficiency. 

The paper is organized as follows.
In Sec.~\ref{sec:method}, we present the modeling framework and methodology adopted in this work.
In Sec.~\ref{sec:result}, we present the results and compare them with observations.
In Sec.~\ref{sec:nu}, we derive the corresponding neutrino spectra from these SNRs.
In Sec.~\ref{sec:dis}, we give a summary and discuss the implications of our results.

\begin{figure}[h]
    \centering
  \includegraphics[width=\linewidth]{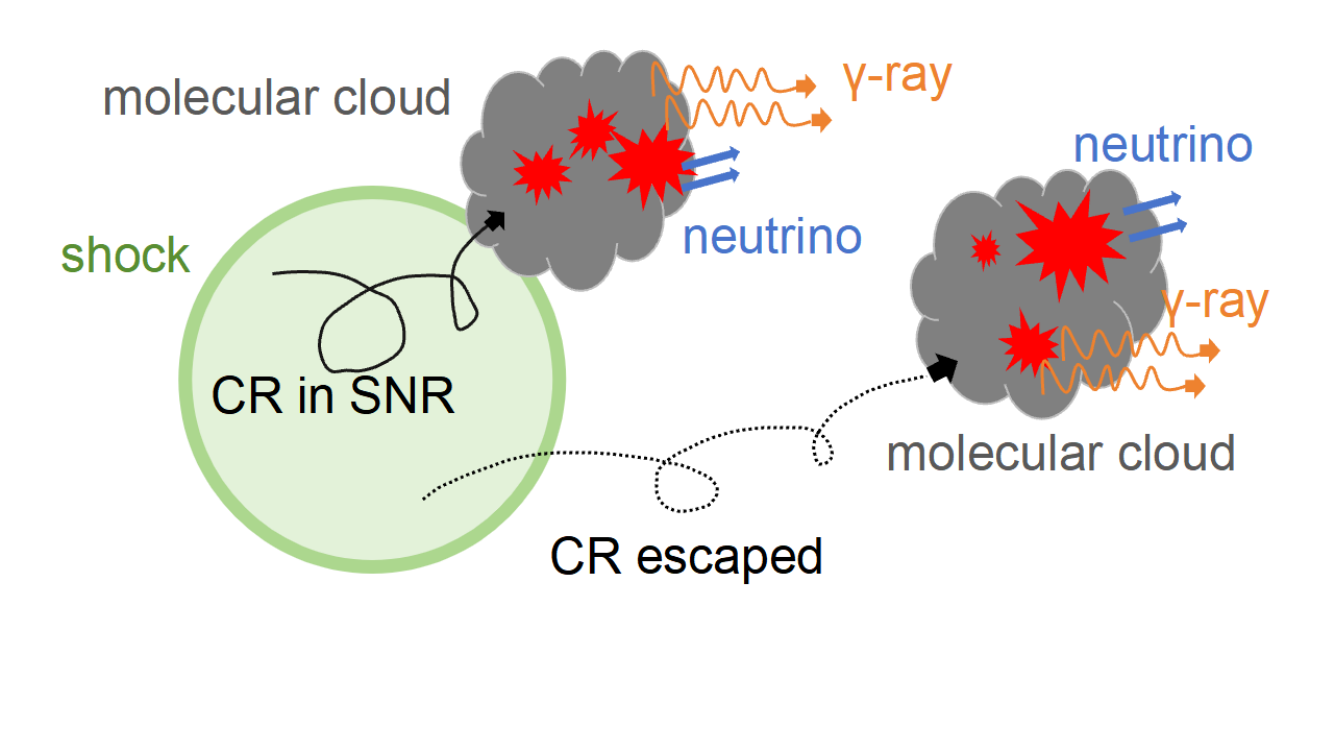}
    \caption{A schematic illustration of the interaction between CRs within the SNR shell and those escaping into the surrounding environment, where they interact with nearby ambient gas clouds. } 
    \label{fig:physical_picture}
\end{figure}

\section{\label{sec:method}Method}
\subsection{Cosmic ray acceleration and escape from SNRs}
Gamma-ray observations of SNRs provide crucial insights into both CR acceleration and escape. The detection of the characteristic ``pion bump'' in the GeV spectra firmly establishes the hadronic origin of the emission, while spatially extended emission beyond the SNR shell provides direct evidence for CR escape. Observations from GeV to PeV thus offer a comprehensive view of CR acceleration at the shock and subsequent escape into the interstellar medium. Despite the wealth of data, a systematic comparison of acceleration and escape parameters across different middle-aged SNRs is still lacking. In particular, it remains unclear how the escape efficiency, diffusion coefficient, and maximum proton energy scale with SNR age and environment. Recent LHAASO observations have extended the spectra of several SNRs to hundreds of TeV, revealing features that challenge the standard DSA picture \cite{cao2026ultrahighenergygammarayimprintspev}.

In this section, we describe the modeling framework used to simulate the acceleration and escape of CRs from SNRs.  
Our model incorporates the time-dependent injection of particles from the SNR shock, assuming spherical symmetry and isotropic diffusion. 
In Fig.~\ref{fig:physical_picture}, we illustrate two scenarios for gamma-ray production from SNRs. 
In the first scenario, CRs inside the SNR shell interact with a molecular cloud swept up by the shock. 
In the second scenario, escaped CRs propagate away from the SNR and interact with a nearby molecular cloud.

Sedov-phase SNRs are middle-aged SNRs in which the swept-up interstellar medium mass is comparable to the ejecta mass, resulting in significant deceleration of the shock wave. 
The dimensionless Sedov onset time is defined as \(t_{\mathrm{ST}}^* \equiv t_{\mathrm{ST}} / t_{\mathrm{ch}}\), 
where the characteristic time scale \(t_{\mathrm{ch}} \equiv \mathcal{E}_{\rm SN}^{-1/2} M_{\rm ej}^{5/6} \rho_0^{-1/3}\) 
is the unique combination of the explosion energy $\mathcal{E}_{\rm SN}$, ejecta mass $M_{\rm ej}$, and ambient density $n_0$. 
With this normalization, the transition time is given by \cite{1999ApJS..120..299T,2018ApJ...866....9L}:
\begin{equation}
t_{\mathrm{ST}}^* \equiv \frac{t_{\mathrm{ST}}}{t_{\mathrm{ch}}}
=
\begin{cases}
0.495 \left[ \frac{5}{3} \frac{3-n}{5-n} \right]^{1/2}, & n < 3, \\
\left[ \frac{2n}{5(n-3)} A^{-5/2} \xi_0^{1/2} \right]^{\frac{2n}{3(n-5)}}, & n > 5,
\end{cases}
\label{eq:sedove_time}
\end{equation}
Here \(n\) is the density power-law index of the ejecta,  $\xi_0 = 2.026$, and \(A\) is a normalization constant that depends on the ejecta structure. 
For \(3 \le n \le 5\), the transition time depends not only on \(n\) but also on the core parameter \(w_{\mathrm{core}} \equiv v_{\mathrm{core}}/v_{\mathrm{ej}}\), 
which characterizes the fraction of mass and energy in the inner core. 
In this regime, no closed-form expression exists; \(t_{\mathrm{ST}}^*\) must 
be obtained by numerically matching the ejecta-dominated and Sedov-Taylor 
blast-wave trajectories for each specific \(w_{\mathrm{core}}\).

The shock acceleration of particles is highly time-dependent: the maximum achievable energy rises with time during the free-expansion phase, peaks at the Sedov phase, and then decays as a power law in time:
\begin{equation}
p_{\text{max},0}(t) = 
\begin{cases} 
p_M \left( \frac{t}{t_{\text{ST}}} \right) & \text{if } t \leq t_{\text{ST}} \\[6pt]
p_M \left( \frac{t}{t_{\text{ST}}} \right)^{-\delta} & \text{if } t > t_{\text{ST}},
\end{cases}
\label{eq:evolution_pM}
\end{equation}
where $p_M$ represents the absolute maximum momentum at the moment when the SNR just enters the Sedov phase. The parameter $\delta$ is a free parameter of the model, a positive number that strongly depends on the time evolution of magnetic turbulence \cite{2019MNRAS.490.4317C}.  
Using Eq.~(\ref{eq:evolution_pM}), we can also define the escape time of particles with a given momentum $p$ as the moment when they can no longer be confined by the SNR and begin to escape. 
The particle escape time is expressed as:
\begin{equation}
t_{\text{esc}}(p) = t_{\text{ST}} \left( \frac{p}{p_M} \right)^{-1/\delta}.
\label{eq:escapetime}
\end{equation}
This means that particles accelerated by the SNR can be divided into two categories according to their escape times: those with $t < t_{\text{esc}}(p)$ are confined particles, and those with $t > t_{\text{esc}}(p)$ are escaping particles. 
It is also useful to define the escape radius:
\begin{equation}
R_{\text{esc}}(p) = R_{\text{sh}}\bigl( t_{\text{esc}}(p) \bigr). 
\label{eq:escape_radius}
\end{equation}

In fact, the particles inside the remnant always originate from the particles accelerated by the shock. 
We assume that the CR pressure at the shock is given by 
$P_{\text{sh}}^{\text{CR}} = \xi_{\text{CR}} \, \rho_0 u_{\text{sh}}^2$, 
where $\xi_{\text{CR}}$ is the fraction of the incoming ram pressure converted into CR pressure.
Therefore, we can obtain the energy spectrum at the shock front
\begin{equation}
f_0(t,p) = \frac{3\xi_{\text{CR}} u_{\text{sh}}^2(t) \rho_0}{4\pi c (m_p c)^4 \Lambda}
\Bigl(\frac{p}{m_p c}\Bigr)^{-\alpha} \Theta\bigl[p_{\text{max},0}(t)-p\bigr],
\label{eq:inject}
\end{equation}
where $m_p$ is the proton mass. 
Here we assume that the particle spectrum at the shock front is a power-law spectrum with a maximum energy. 
The slope $\alpha$ is a free parameter, although DSA predicts $\alpha = 4$ for strong shocks \cite{Drury_1983}. 
To ensure that the integrated CR pressure at the shock matches 
$P_{\mathrm{sh}}^{\mathrm{CR}} = \xi_{\mathrm{CR}}\rho_0 u_{\mathrm{sh}}^2$, 
we introduce the normalization constant $\Lambda$ as
\begin{equation}\label{eq:nondimensionalize}
\Lambda = \int_{p_{\text{min}}/(m_p c)}^{p_{\text{max},0}/(m_p c)} 
y^{4-\alpha} (1+y^2)^{-1/2} \, dy,
\end{equation}
where $y \equiv p/(m_p c)$ is the dimensionless momentum. 
The integrand accounts for both the phase-space volume 
($y^{4-\alpha}$) and the relativistic velocity correction 
($(1+y^2)^{-1/2}$), which are crucial for an accurate 
evaluation of the CR energy budget over the entire 
momentum range—from non-relativistic to ultra-relativistic 
particles.
The evolution of the phase-space density $f(t,r,p)$ for accelerated CRs in spherical symmetry is described by the following transport equation:
\begin{equation}
\frac{\partial f}{\partial t} + u \frac{\partial f}{\partial r} = \frac{1}{r^2} \frac{\partial}{\partial r} \left( r^2 D \frac{\partial f}{\partial r} \right) + \frac{1}{r^2} \frac{\partial (r^2 u)}{\partial r} \frac{p}{3} \frac{\partial f}{\partial p}, 
\label{eq:Convection-Diffusion Equation}
\end{equation}
where $u(t,r)$ denotes the plasma advection velocity and $D(t,r,p)$ is the spatial diffusion coefficient.

Particles with $t < t_{\text{esc}}(p)$, are tightly confined by the strong turbulence at the shock, and their spatial distribution is mainly determined by the shock position and expansion history, rather than by the diffusion process. Therefore, the diffusion term can be neglected compared to the advection and adiabatic terms. The evolution of confined particles can then be expressed as follows~\cite{2019MNRAS.490.4317C},
\begin{equation}
\frac{\partial f_{\text{conf}}}{\partial t} + u \frac{\partial f_{\text{conf}}}{\partial r} = \frac{1}{r^2} \frac{\partial (r^2 u)}{\partial r} \frac{p}{3} \frac{\partial f_{\text{conf}}}{\partial p}.
\label{eq:Fokker-Planck}
\end{equation}
For $u(t,r)$, we employ the linear velocity approximation, which assumes that for $r \leq R_{\text{sh}}$ the plasma velocity profile takes the form \cite{RevModPhys.60.1}
\begin{equation}
u(t,r) = \left(1 - \frac{1}{\sigma}\right) \frac{u_{\text{sh}}(t)}{R_{\text{sh}}(t)} \, r.
\label{eq:plasma_velocity}
\end{equation}
Here, $\sigma$ represents the compression ratio at the shock. For strong shocks, $\sigma = 4$. The solution for the particle spectrum is expressed as
\begin{equation}
f_{\text{conf}}(t, r, p) = f_0(p, t) \left( \frac{t'}{t} \right)^{\epsilon} \frac{\Lambda(t)}{\Lambda(t')} \Theta[p_{\text{max}}(t, r) - p]. 
\label{eq:confined_particle}
\end{equation}
The quantity $t'(t,r)$ is defined as the time at which the plasma layer now located at radius $r$ at time $t$ was first swept by the shock. The function $p_{\text{max}}(t, r)$ denotes the maximum momentum of particles located at position $r$ and time $t$. It is obtained by taking the maximum momentum of particles that were accelerated at the earlier time $t'$ and then reducing their energy through adiabatic losses that occur between $t'$ and $t$. In other words,
\begin{equation}
p_{\text{max}}(t, r) = p_{\text{max},0}(t) \left( \frac{t'}{t} \right)^{\frac{2(\sigma-1)}{5\sigma} - \delta}. 
\label{eq:pM}
\end{equation}

For escaping particles, the model assumes that when $t > t_{\text{esc}}(p)$, particles with momentum $p$ can no longer be confined by the turbulence in the shock region and begin to escape. However, escaping particles do not leave the SNR directly; they also propagate inside the remnant for a certain period of time, and even particles that have escaped the SNR have a finite probability of returning to the interior of the remnant. In fact, assuming that the behavior of escaping particles is almost independent of the shock, we neglect the advection and adiabatic terms and obtain the distribution of escaping particles~\cite{2019MNRAS.490.4317C},
\begin{equation}
\frac{\partial f_{\text{esc}}}{\partial t} = \frac{1}{r^2} \frac{\partial}{\partial r} \left[ r^2 D(p) \frac{\partial f_{\text{esc}}}{\partial r} \right], 
\label{eq:diffusion_equation}
\end{equation}
The equation is to be solved with initial conditions at time $t = t_{\text{esc}}(p)$, where the distribution function of confined particles is defined as $f_{\text{conf}}(t_{\text{esc}}(p), r, p) \equiv f_{\text{conf},0}(r, p)$. The initial condition for the escaping particles is:
\begin{equation}
f_{\text{esc}}(t_{\text{esc}}(p), r, p) = 
\begin{cases}
f_{\text{conf},0}(r, p), & r < R_{\text{sh}}\bigl(t_{\text{esc}}(p)\bigr), \\[4pt]
0, & \text{elsewhere}.
\end{cases}
\label{eq:initial_condition}
\end{equation}
For the diffusion coefficient, we assume a relation with the average Galactic diffusion coefficient
\begin{equation}
D_{\text{out}}(p) \equiv \chi D_{\text{Gal}}(p) = \chi 10^{28} \left( \frac{pc}{10 \, \text{GeV}} \right)^{1/3} \, \text{cm}^2 \, \text{s}^{-1}, 
\label{eq:diffusion_coefficient}
\end{equation}
where $\chi$ is a constant representing the ratio between the actual diffusion coefficient and the Galactic average. 
Furthermore, the diffusion coefficient is assumed to be the same inside and outside the remnant. The solution of Eq.~(\ref{eq:diffusion_equation}) has an analytic form only for certain values of $\alpha$ \cite{2019MNRAS.490.4317C}. 
Introducing the diffusion time
\begin{equation}
\Delta t \equiv t-t_{\rm esc}(p),
\end{equation}
the corresponding Green's function is
\begin{equation}
G(\mathbf{r},\mathbf{r}',\Delta t)
=
\frac{1}
{\left(4\pi D(p)\Delta t\right)^{3/2}}
\exp\!\left[
-\frac{|\mathbf{r}-\mathbf{r}'|^2}
{4D(p)\Delta t}
\right].
\end{equation}
Therefore, the escaping particle distribution can be written generally as
\begin{equation}
f_{\rm esc}(\mathbf{r},t,p)
=
\Theta\!\left[t-t_{\rm esc}(p)\right]
\int
G(\mathbf{r},\mathbf{r}',\Delta t)
f_{\rm conf,0}(\mathbf{r}',p)
\,d^3r',
\label{eq:green_solution}
\end{equation}
where $f_{\rm conf,0}$ denotes the confined particle distribution at the escape time.

Assuming spherical symmetry, Eq.~(\ref{eq:green_solution}) reduces to the one-dimensional integral
\begin{equation}
\begin{split}
f_{\rm esc}(r,t,p)
=
&
\frac{\Theta\!\left[t-t_{\rm esc}(p)\right]}
{\sqrt{\pi}R_d\,r}
\\
&
\times
\int_0^{R_{\rm sh}}
r'
f_{\rm conf,0}(r',p)
\\
&
\times 
\left[
e^{-(r-r')^2/R_d^2}
-
e^{-(r+r')^2/R_d^2}
\right]
dr',
\end{split}
\label{eq:escape_particle}
\end{equation}
where
\begin{equation}
R_d=\sqrt{4D(p)\left[t-t_{\rm esc}(p)\right]}.
\label{eq:RD}
\end{equation}
In fact, having determined the distributions of confined and escaping particles at any time and any position, we can obtain the spatial distribution of particles for any energy. 

Therefore, we can derive the particle distribution inside the remnant shell at the current age:
\begin{equation}
J_p^{\text{in}}(t, p) = \frac{4\pi}{V_{\text{SNR}}} \int_0^{R_{\text{sh}}(t)} \bigl[ f_{\text{esc}}(t, r, p)  + f_{\text{conf}}(t, r, p) \bigr] r^2 dr, 
\label{eq:shell_modle}
\end{equation}
where $V_{\text{SNR}} = \frac{4}{3}\pi R_{\text{sh}}^3(t)$ is the remnant volume.
When the SNR shock interacts with a molecular cloud, it can produce radiation that lights up the shell. 

On the other hand, we can derive the particle distribution at any position outside the remnant, which consists solely of escaping particles, 
\begin{equation}
J_p^{\text{out}}(t, p) = \frac{3}{R_2^3 - R_1^3} \int_{R_1}^{R_2} \bigl[ f_{\text{esc}}(t, r, p)   \bigr] \, r^2 \, \mathrm{d}r.
\label{eq:escape_model}
\end{equation}

\subsection{Production of high-energy gamma rays}
High-energy gamma-rays are produced through proton-proton ($pp$) interactions via the decay of neutral pions.
The expected gamma-ray flux observed at Earth from a molecular cloud illuminated by CRs can be expressed as \cite{2020A&A...635A..40P}:
\begin{equation}
\phi(E_\gamma) = \frac{M_{\text{cl}}}{4\pi d_l^2 m_{\text{avg}}} \int_{T_p^{\text{min}}}^{T_p^{\text{max}}} 4\pi J_p(T_p) \varepsilon(T_p) \frac{d\sigma_{pp}(T_p, E_\gamma)}{dE_\gamma} dT_p, 
\label{eq:gamma_ray}
\end{equation}
In the above expression, $J_p(T_p) = (v/4\pi)\,n_p(T_p)$ denotes the CR proton intensity as a function of the kinetic energy $T_p$ (where $v$ is the particle velocity). The quantity $M_{\text{cl}}$ represents the total mass of the molecular cloud, and $m_{\text{avg}}$ is the average atomic mass of the cloud gas, which, for solar abundances, is $m_{\text{avg}}\approx1.4\,m_p$. The upper integration limit $T_p^{\text{max}}$ is the maximum kinetic energy of the accelerated protons. The lower limit $T_p^{\text{min}}$ corresponds to the threshold kinetic energy for $\pi^0$ production in $pp$ collisions. The term $d\sigma_{pp}(T_p,E_\gamma)/dE_\gamma$ is the differential cross section for gamma-ray production. The nuclear enhancement factor $\varepsilon(T_p)$, which accounts for gamma rays from nucleus-nucleus interactions, is set to 1.5 \cite{1997ApJ...478..225M,Koldobskiy_2021,Kamae_2006,Kachelrie__2019,Kafexhiu_2014}.

\section{\label{sec:result} Results}
In this section, we model the gamma-ray emission from four archetypal middle-aged SNRs: W51C, IC~443, W44, and W28. 
The four selected SNRs represent some of the best studied cases of SNR–molecular cloud interactions and offer complementary insights into different aspects of CR acceleration and escape~\cite{Ohira_2010}. 
The core advantage of this model is its capability to simultaneously describe the hadronic emission from the remnant interior and the extended emission produced by escaped particles interacting with molecular clouds outside, thereby allowing us to impose observational constraints on parameters such as the diffusion coefficient, maximum proton energy, and spectral cutoff \cite{2019MNRAS.490.4317C}.

To constrain the model parameters, we perform a Bayesian inference using the Markov Chain Monte Carlo (MCMC) technique. We employ the affine-invariant ensemble sampler implemented in the open-source Python package \texttt{emcee}~\citep{Foreman_Mackey_2013}.
The log-likelihood function is constructed assuming Gaussian errors on the observed flux points:
\begin{equation}
\ln \mathcal{L}(\boldsymbol{\theta}) = -\frac{1}{2} \sum_{i} \frac{\left[ F_{\mathrm{mod}}(E_i; \boldsymbol{\theta}) - F_{\mathrm{obs}}(E_i) \right]^2}{\sigma_i^2} \;,
\end{equation}
where \(\boldsymbol{\theta}\) denotes the set of free model parameters, \(F_{\mathrm{mod}}\) and \(F_{\mathrm{obs}}\) are the model-predicted and observed gamma-ray fluxes at energy \(E_i\), respectively, and \(\sigma_i\) are the corresponding \(1\sigma\) measurement uncertainties.
We run the MCMC sampling with $N_{\mathrm{walkers}} = 20$ walkers for a total of $20\,000$ steps. The first $1\,000$ steps are discarded as burn-in to ensure that the chains have reached a stationary state.
The \(1\sigma\) uncertainties are derived from the 16th and 84th percentiles.
The gamma-ray data used in the fits are collected from published GeV-TeV measurements of the four remnants \cite{2012A&A...541A..13A,2012A&A...541A..13A,2024SciBu..69.2833C,2026PhRvL.136p1002C,2025A&A...693A.255A,2008A&A...481..401A,Cui_2018}.

\subsection{W51C}
W51C is an SNR identified by its radio shell, located at a distance of approximately 5.5 kpc from Earth, with an age estimated to be between 10,000 and 30,000 years \cite{2008xmm..prop...96L,2009ApJ...706L...1A,2022MNRAS.516.6055R,1995ApJ...447..211K,2010ApJ...720.1055S}, and an initial supernova explosion energy of about $3.6 \times 10^{51}$~erg \cite{1995BKAS...20....5K}.
W51C belongs to the larger W51 complex, which is one of the largest star-forming regions in our Galaxy and is divided into two parts, W51A and W51B.
The SNR is located at the south-eastern extremity of W51B, and the interaction between the SNR shell and the MC W51B is evidenced by the detection of two OH (1720 MHz) masers \cite{1997AJ....114.2058G}.
Ref.~\cite{2016ApJ...816..100J} used Fermi-LAT data to study the spectrum above 60 MeV of the middle-aged SNR W51C. 
Their results show a clear break in the power-law gamma-ray spectrum at 290 MeV and this break is most likely associated with the energy production threshold of $\pi^0$ mesons. 
The gamma-ray spectrum shows a high‑energy break at energy around 2.7 GeV,
and is in good agreement with the MAGIC data for energies above 75 GeV \cite{2012A&A...541A..13A}. 
LHAASO has, for the first time, measured gamma-ray emission from the W51 complex beyond 100~TeV and detected a clear spectral break around tens of TeV \cite{2024SciBu..69.2833C}. 
These results strongly indicate that extreme CR accelerators exist within the W51 complex. 

\begin{figure}[h]
    \centering
    \includegraphics[width=\linewidth]{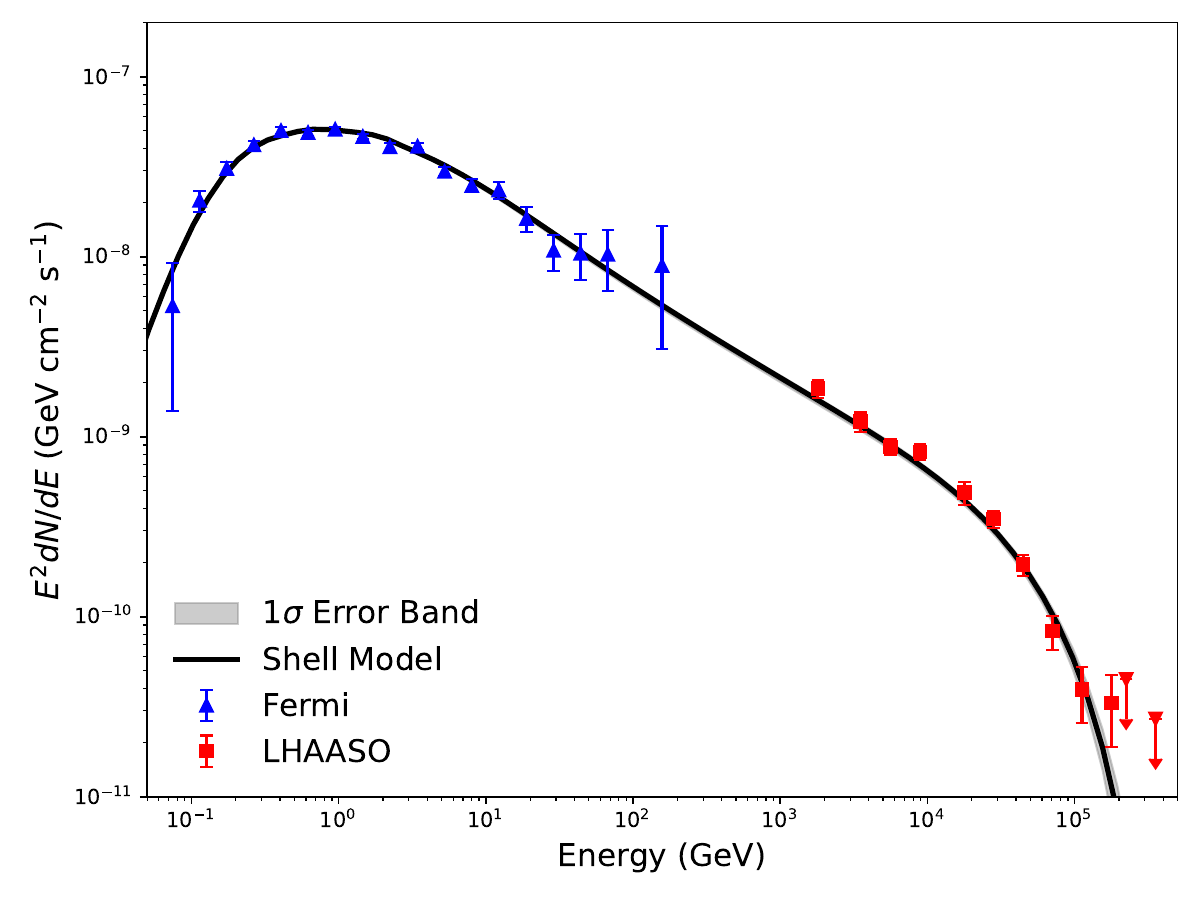}    
    \includegraphics[width=\linewidth]{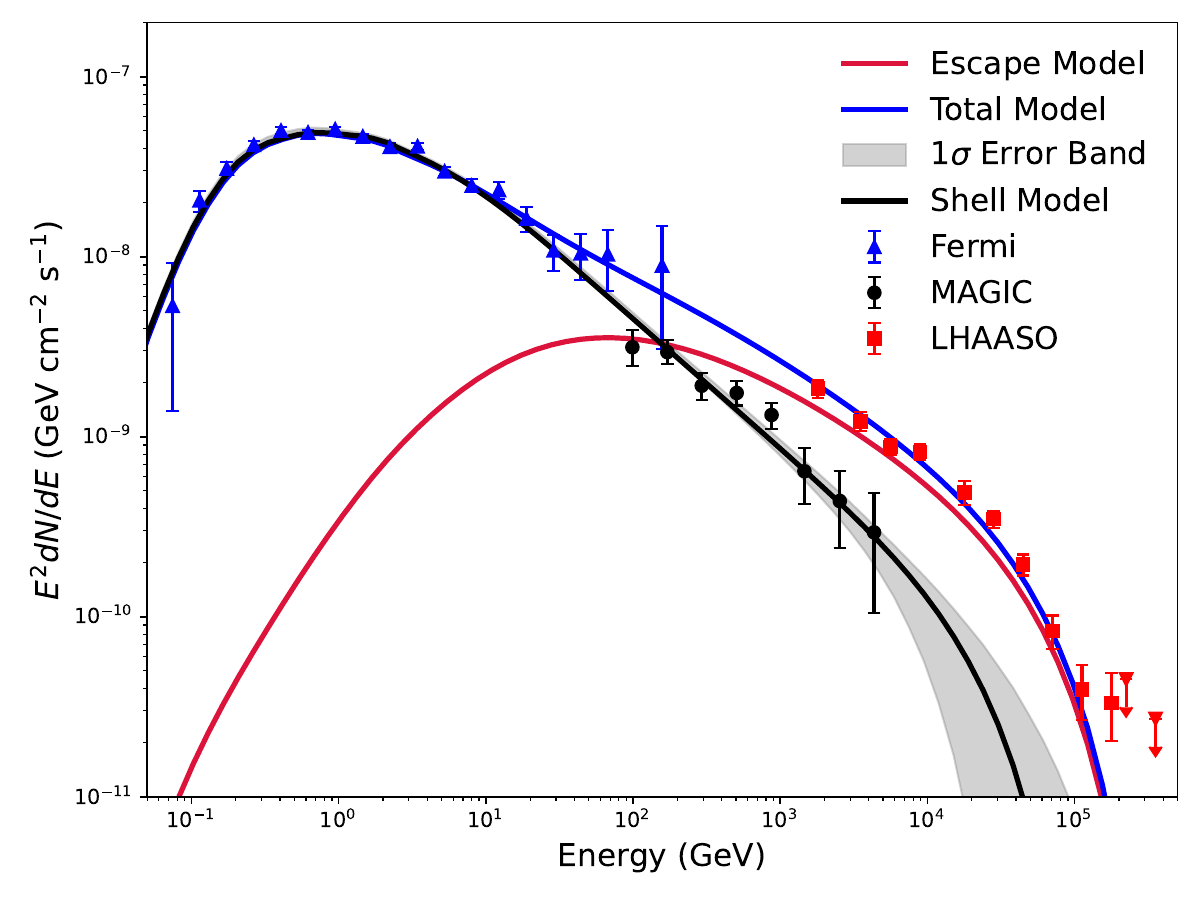}
    \caption{Modeled gamma-ray spectra of W51C compared with observations from scenario 1 (upper panel) and scenario 2 (lower panel), respectively.
}
    \label{fig:w51}
\end{figure}

For W51C, we adopt a current shell radius of \(24\ {\rm pc}\). 
Our results are shown in Fig.~\ref{fig:w51}. 
We explore two scenarios: (1) the gamma-ray emission measured by Fermi-LAT and LHAASO originates from a common component, namely CRs within the shell interacting with shock‑swept molecular clouds; (2) the gamma-ray  emission measured by Fermi-LAT and MAGIC arises from the shell component, whereas the LHAASO emission originates from escaped CRs.

In the upper panel of Fig.~\ref{fig:w51}, we show the fitting results for scenario 1. The current maximum acceleration energy of the SNR is estimated to be \(\sim 1.5\) GeV, assuming an injected maximum energy of $\sim$\(540\) TeV at the Sedov phase and a temporal decay of the maximum acceleration energy as $t^{-\delta}$ with $\delta \sim 3.9$.
The energy conversion efficiency required to produce the observed gamma-ray emission from hadronic interactions between the CRs and the cloud is estimated to be $\eta_{\rm cr} = 4\%$ for a molecular cloud of mass $1.9\times10^5\,M_\odot$.
For comparison, Ref.~\cite{2024SciBu..69.2833C} derived a total CR energy $W_{\rm cr} \simeq 1.3\times10^{50}$~erg for W51C (for $d = 5.5$~kpc and $n_H = 100$~cm$^{-3}$), corresponding to $\eta_{\rm cr} \simeq 3.6\%$ relative to $\mathcal{E}_{\rm SN} = 3.6\times10^{51}$~erg. However, this efficiency should be interpreted with caution, as $W_{\rm cr}$ scales inversely with the target gas density ($W_{\rm cr} \propto n_H^{-1}$). A lower ambient gas density would require a larger $W_{\rm cr}$ to reproduce the same gamma-ray flux, thus increasing the inferred efficiency. 

In the lower panel of Fig.~\ref{fig:w51}, we show the results for scenario~2, where the LHAASO data are interpreted as emission from CRs that have escaped the SNR shell and interact with a nearby molecular cloud. To reproduce the observed flux, a molecular cloud of mass \(1.2\times10^5\,M_\odot\) located \(\sim58\) pc from the SNR and a maximum injection energy of \(500\) TeV are required. 
The TeV emission measured by MAGIC is consistent with CR interactions in the shell region \cite{2012A&A...541A..13A}.
Ref.~\cite{Carpenter_1998} identified the interacting $68\ {\rm km\,s^{-1}}$ molecular cloud with a total mass of $1.9\times10^{5}\,M_{\odot}$, but MAGIC observations constrain the actively emitting volume—radius $\sim24$ pc, density $n_{\rm H}\sim10\ {\rm cm^{-3}}$—to a mass of $2.1\times10^{4}\,M_{\odot}$, or $\sim11\%$ of the total cloud mass  \cite{2012A&A...541A..13A}. In this case, the inferred CR acceleration efficiency is $\eta_{\rm cr}\sim 11\%$.

As shown in Table~\ref{tab:params}, the spectral indices and maximum energies of accelerated CRs differ between the two scenarios. Scenario 1 requires a hard injection spectrum with \(\alpha \sim 4\) and a high maximum acceleration energy of \(p_M \sim 500\) TeV, which is rather extreme for typical CR acceleration processes in SNRs. In contrast, scenario 2, in which the LHAASO emission originates from escaped CRs interacting with nearby molecular clouds, provides a more natural interpretation of the observed UHE emission.
Recently, Ref.~\cite{Sunny:2026jnz} showed that the shock-cloud interaction scenario of W51C failed to explain both the gamma-rays from Fermi-LAT and LHAASO.
This is consistent with the findings in this work, where we find that an injection spectrum with $\alpha \sim 4$ is required.
This value is essentially consistent with the test-particle diffusive shock acceleration (DSA) prediction of \(\alpha = 4\), but is notably harder than the value of $\alpha = 4.4$ adopted in Ref.~\cite{Sunny:2026jnz}.

\subsection{IC 443}
IC 443 is a middle-aged SNR at a distance of about 1.5 kpc \cite{1984ApJ...281..658F}, with its age estimated to be between 3,000 and 30,000 years \cite{2008A&A...485..777T,2008AJ....135..796L,1988ApJ...335..215P}. 
The OH maser emission and various molecular lines from IC 443 make it the most evident case of a supernova interacting with a molecular cloud~\cite{1999ApJ...511..798C,2006ApJ...652.1288H,2008ApJ...683..189H,1986ApJ...302L..63H,1998ApJ...505..286S,2014ApJ...788..122S}. 
For IC 443, the estimated supernova explosion energy $\mathcal{E}_{\text{SN}}$ typically ranges from $1$ to $2.5 \times 10^{51}$~erg, with a commonly adopted reference value of $1 \times 10^{51}$~erg \cite{2021A&A...649A..14U,2013Sci...339..807A}. 
The hadronic origin of the gamma-ray emission from IC 443 is firmly established by the detection of a pion-decay feature in the low-energy spectrum by Fermi-LAT \cite{2010ApJ...712..459A,2013Sci...339..807A}, with a broken power-law proton spectrum providing an excellent fit to the broadband data. 
LHAASO observations of IC 443 have, for the first time, resolved two gamma-ray components in the TeV band: a point source C0 and an extended source C1~\cite{2026PhRvL.136p1002C}. 
C0 coincides spatially with the compact source detected by Fermi-LAT, which exhibits the spectral signature of $\pi^0$ decay in the low-energy spectrum. 
C1 is an extended source that may be associated with the interactions of escaped CRs from IC~443, hadronic emission from an older SNR G189.6+3.3, or leptonic emission from a pulsar wind nebula ~\cite{2026PhRvL.136p1002C}.

\begin{figure}[!h]
    \centering
    \includegraphics[width=\linewidth]{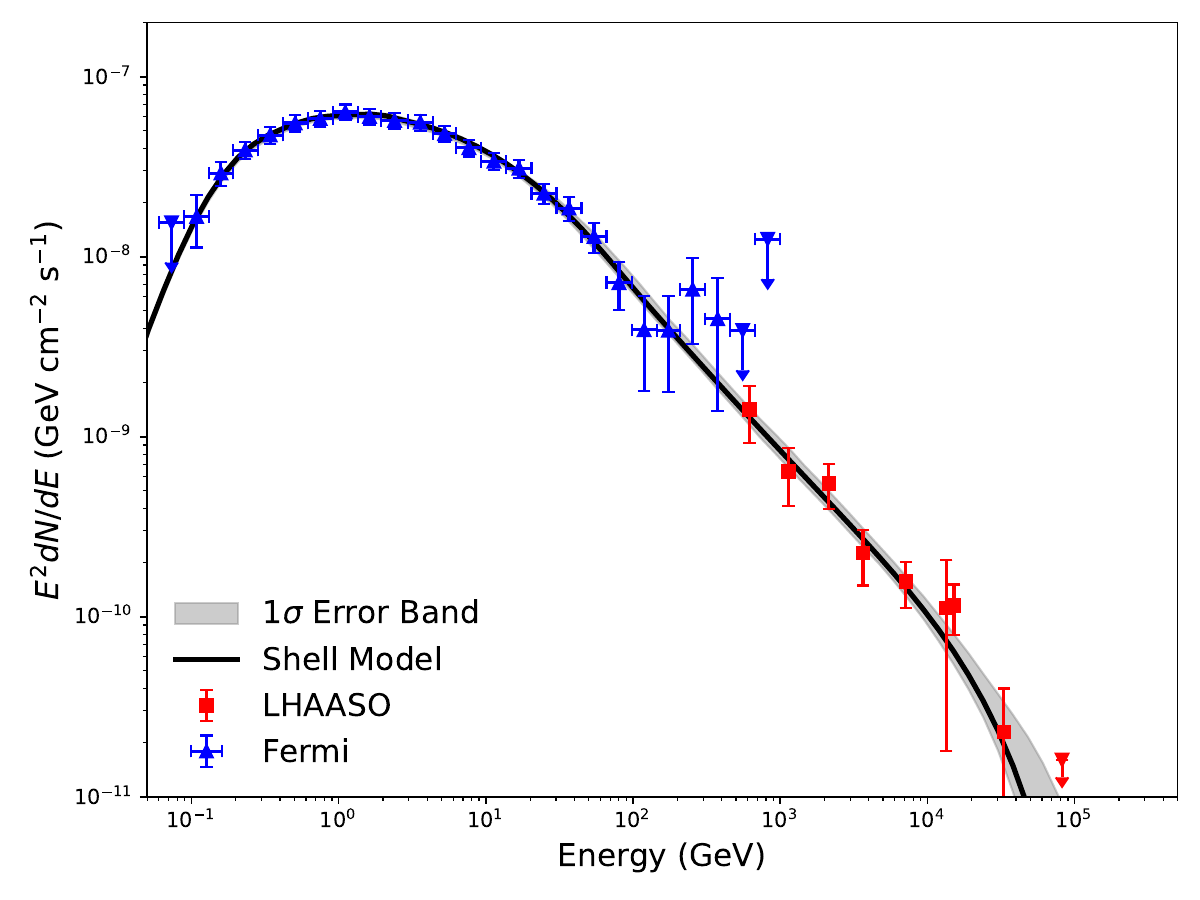}
    \includegraphics[width=\linewidth]{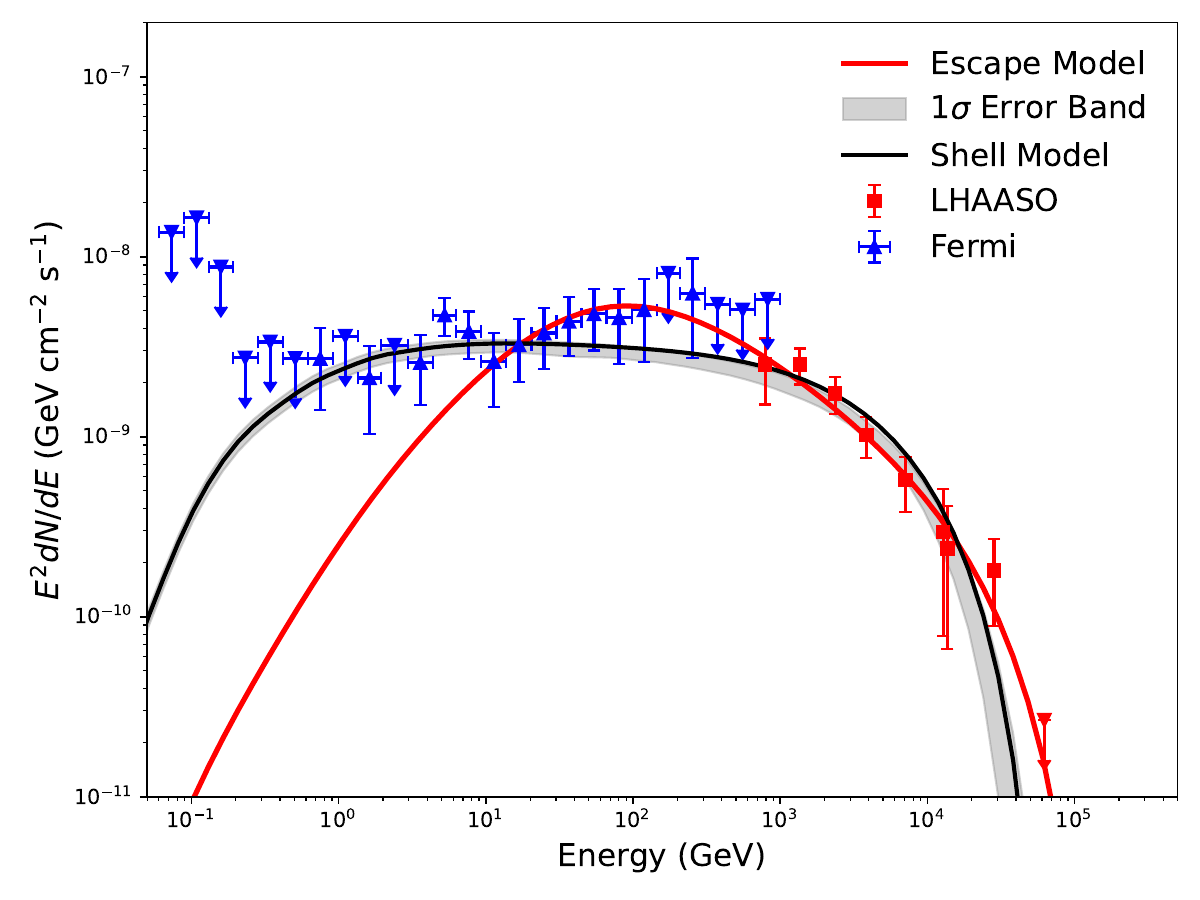}
    \caption{Similar to Fig.~\ref{fig:w51}, the modeled gamma-ray emission from CRs inside the shell compared with the C0 component (upper panel). 
The lower panel shows the interpretation of the C1 component as gamma-ray emission from escaped CRs interacting with a nearby molecular cloud.}
    \label{fig:ic443}
\end{figure}

For IC 443, we adopt a current shell radius of \(10\ {\rm pc}\) in our modeling. The fitting results are presented in Fig.~\ref{fig:ic443}. Our analysis showed that the gamma-ray spectrum of the C0 component, covering Fermi-LAT and LHAASO observations, can be adequately reproduced by CRs within the SNR shell. 
Assuming a molecular cloud mass of \(10^3\,M_\odot\) \cite{2013Sci...339..807A}, we derive a CR acceleration efficiency of \(\sim 4\%\).
The current maximum proton energy is estimated to be \(\sim 121\) GeV for an injected maximum energy of \(220\) TeV that decreases with an index $\delta \sim 3$.
For the C1 component, we attribute the observed emission to the interaction between escaped CRs and a molecular cloud with a mass of $6.8\times 10^3\,M_\odot$ at a distance of $18\,\text{pc}$ from the SNR, as shown in Fig.~\ref{fig:ic443}. However, the escaped-CR model provides a poorer fit below 10 GeV.

We additionally investigate the possibility that the C1 component is produced by the shell emission of G189.6+3.3, an old SNR with an age of about 100,000 years \cite{1994A&A...284..573A}.
The results are shown in the lower panel in Fig.~\ref{fig:ic443}.
Assuming that the remnant is currently interacting with a molecular cloud of mass \(10^3\,M_\odot\), we derive an energy conversion efficiency of about \(2\%\), see Table~\ref{tab:params}. 
Our modeling of the C1 component under the assumption that it originates from the shell emission of G189.6+3.3 yields a spectral index of \(\alpha \simeq 4.03\) for the accelerated protons. 
Furthermore, our results indicate that particle acceleration in G189.6+3.3 has ceased at the present epoch, consistent with its old age (\(\sim 10^5\) yr) and its classification as a remnant that has entered the radiative phase of evolution. This is in line with the picture that old SNRs like G189.6+3.3 no longer confine or accelerate particles efficiently, and any observed gamma-ray emission from such systems would be dominated by pre-accelerated particles that have escaped.
In addition, the diffusion coefficient for 10~GeV protons in the surrounding medium is estimated to be $\sim 2\times 10^{24}~\text{cm}^{2}\,\text{s}^{-1}$, which is about four orders of magnitude below the Galactic average. 
In summary, the shell-emission interpretation of the C1 component from G189.6+3.3 requires extreme parameter values.

\subsection{W44}
The SNR W44 serves as an ideal laboratory for studying CR acceleration and escape processes. 
Its initial explosion energy $\mathcal{E}_{\text{SN}}$ is commonly assumed to be on the order of $5\times10^{51}$~erg \cite{2013Sci...339..807A}.
W44 is a middle‑aged mixed‑morphology SNR (age $\sim$ 20~kyr, distance $\sim$ 2.2kpc) that exhibits a well‑defined radio shell and bright thermal X-ray emission from the interior. 
The interaction between the SNR shell and the molecular cloud is evidenced by the detection of OH (1720 MHz) masers \cite{1997ApJ...489..143C}. 
Its associated pulsar, PSR B1853+01, has a spin‑down age consistent with the remnant's evolutionary stage \cite{1996ApJ...464L.165F}. 
Observations by Fermi‑LAT \cite{2020ApJ...896L..23P,2025A&A...693A.255A} detected the $\pi^0$-decay ``pion bump'' in the GeV spectrum, firmly establishing that the gamma‑ray emission originates from hadronic interactions of accelerated CR protons with dense molecular gas.

\begin{figure}[h]
    \centering
    \includegraphics[width=1.\linewidth]{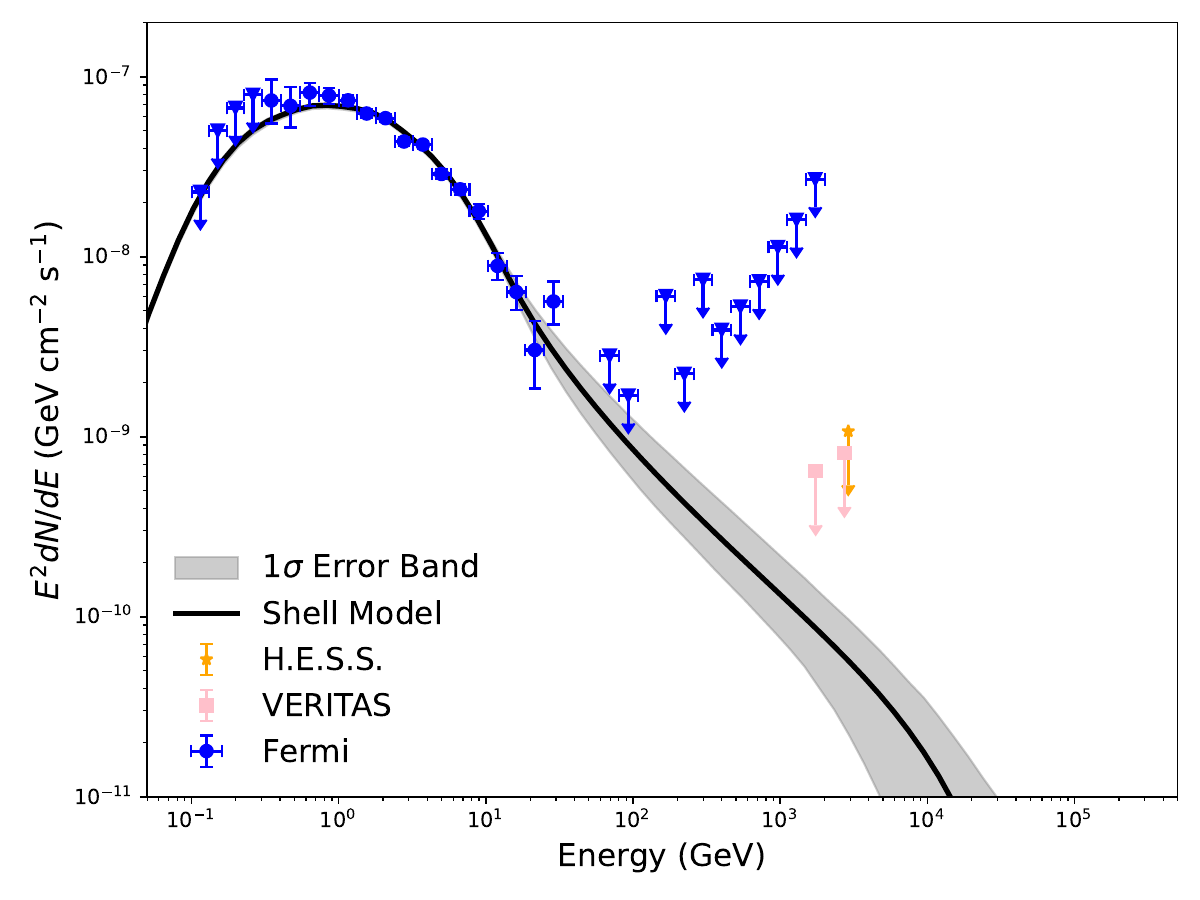}
    \includegraphics[width=1.\linewidth]{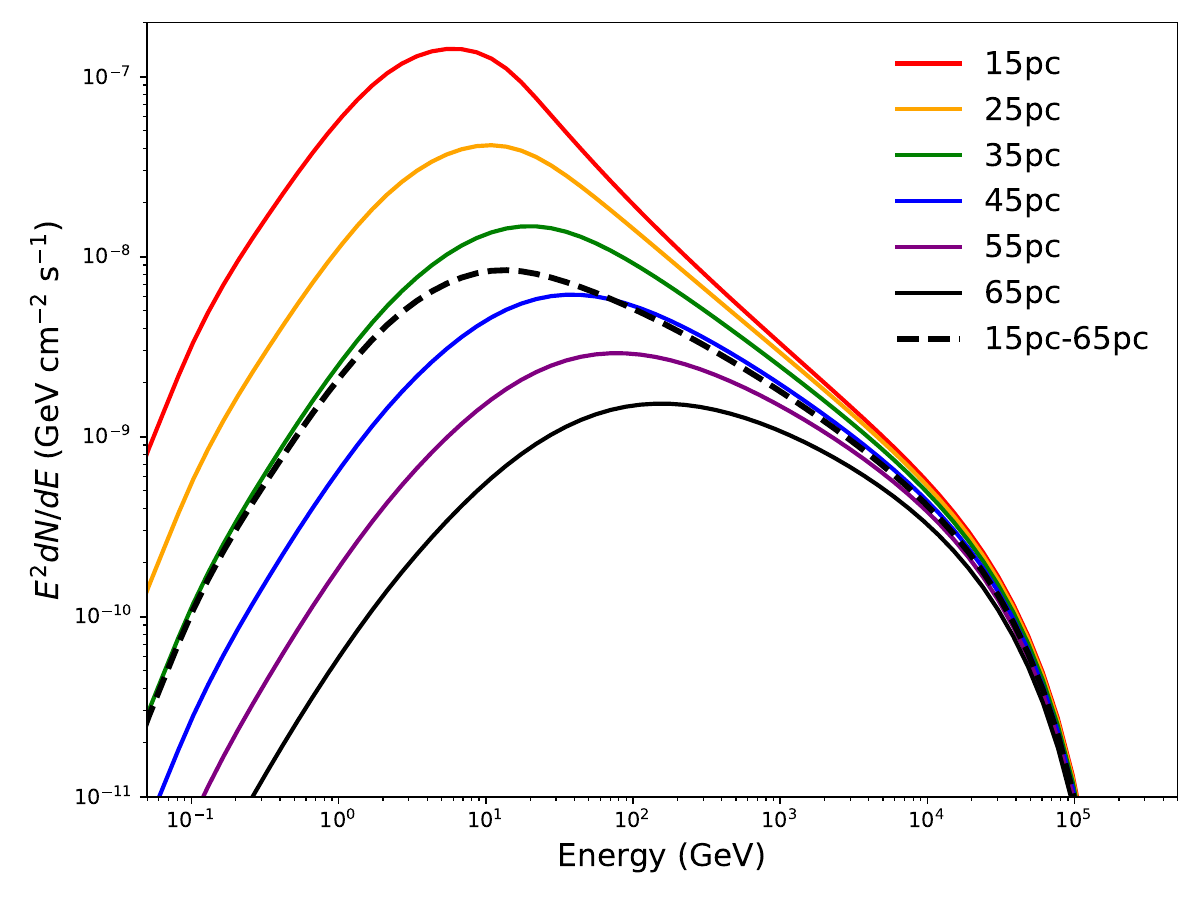}
    \caption{Similar to Fig.~\ref{fig:w51}, we compare the modeled gamma-ray emission from CRs within the SNR shell with the Fermi-LAT data and the upper limits from VERITAS and H.E.S.S. In the lower panel, we present the theoretical expectation for gamma-ray emission produced by escaped CRs interacting with molecular clouds located at different distances from the SNR.}
    \label{fig:w44}
\end{figure}

For W44, we adopt a current shell radius of \(13\ {\rm pc}\). As shown in Fig.~\ref{fig:w44}, we model the gamma-ray emission as arising from hadronic \(pp\) interactions between accelerated CRs and a molecular cloud of mass \(5\times10^3\,M_\odot\) \cite{2013Sci...339..807A}, and the derived acceleration efficiency is \(\eta_{\rm cr} \sim 1\%\). 
For an injected maximum energy of \(105\) TeV at the beginning of the Sedov phase that decreases with time \(t^{-1.82}\), the current maximum proton energy is estimated to be \(\sim 28\) GeV.

We also present the gamma-ray spectra resulting from the interactions of escaped CRs with a giant molecular complex with a mass \(1.24\times10^5\,M_\odot\)~\cite{2020ApJ...896L..23P} surrounding W44 in Fig.~\ref{fig:w44}. 
The dashed black line shows the gamma-ray spectrum integrated over a uniform spatial distribution of the molecular complex spanning distances of 15–65 pc from W44.

\subsection{W28}
SNR W28 is a prototypical thermal composite SNR, at a distance between 1.8 and 3.3 kpc \cite{1976Ap&SS..40...91G,1981SvAL....7...17L} and an age between 35,000 and 150,000 yr \cite{1993ApJ...409L..57K}. 
Although detailed modeling of W28 varies, the theoretically estimated range of its initial explosion energy is generally accepted to be of the order of the canonical supernova explosion energy, $\sim 1 \times 10^{51}$~erg \cite{2002AJ....124.2145V,2020A&A...635A..40P}. 
There are four molecular clouds around W28, namely MCs N, A, B, and C. 
In fact, the SNR is currently interacting with MCs N, as traced by the detection of OH (1720 MHz) masers \cite{1994ApJ...424L.111F,1999ApJ...522..349C,1997ApJ...489..143C}. 
H.E.S.S. observations of W28 have identified four TeV gamma-ray sources: HESS J1801-233, situated along the remnant's northeastern edge, and a triplet of sources (HESS J1800-240A, B, and C) located to the south, beyond the radio boundary~\cite{2008A&A...481..401A}. 
The spatial coincidence between the southern H.E.S.S. sources and molecular clouds at distances consistent with that of W28 suggests that these sources could be produced by escaped CRs from the SNR \cite{10.1093/mnras/sts450,Cui_2018}. 
The gamma-ray emission from HESS J1800-240C (associated with MCs C) is thought to be contaminated by the SNR candidate G5.71-0.08 \cite{2008A&A...481..401A,Hanabata_2014}. 

\begin{figure}[!hb]
    \centering
    \includegraphics[width=1.\linewidth]{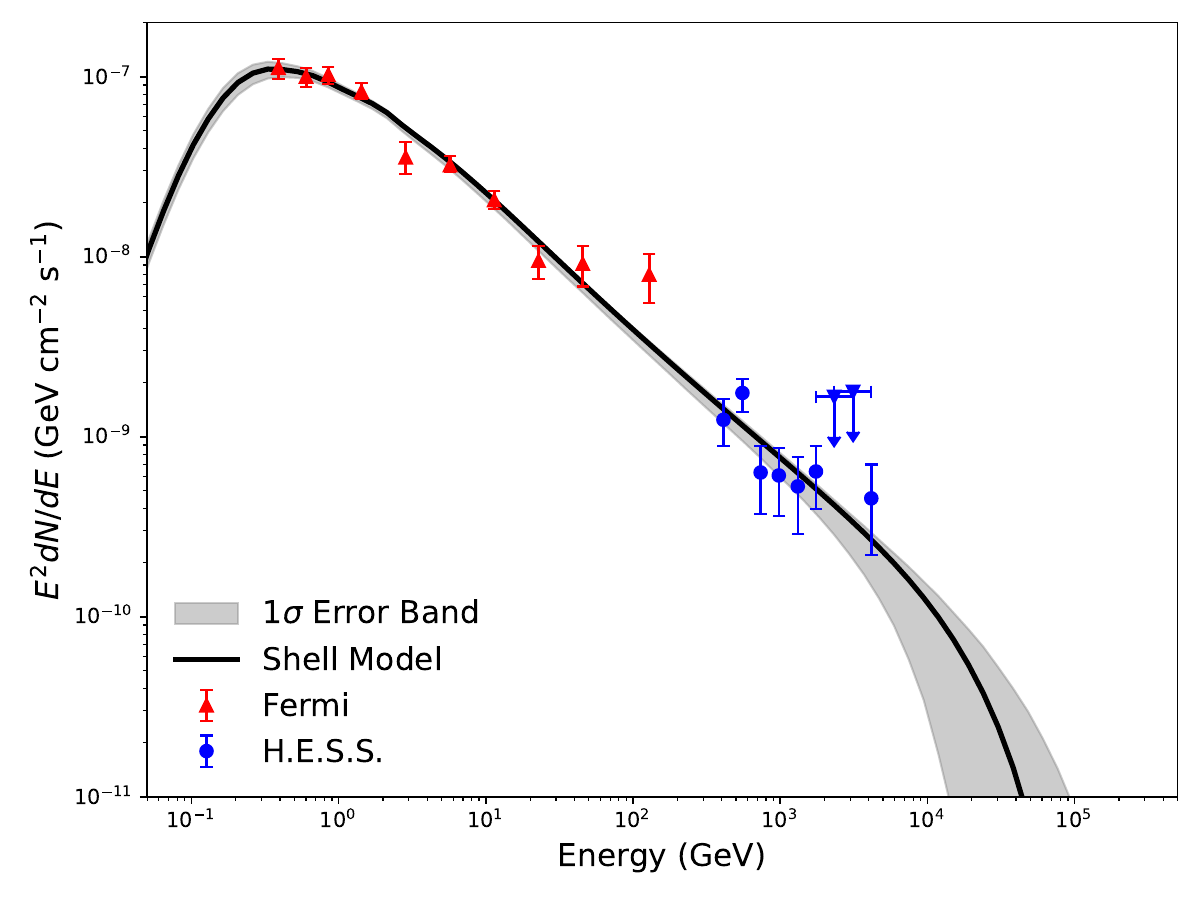}
    \includegraphics[width=1.\linewidth]{
    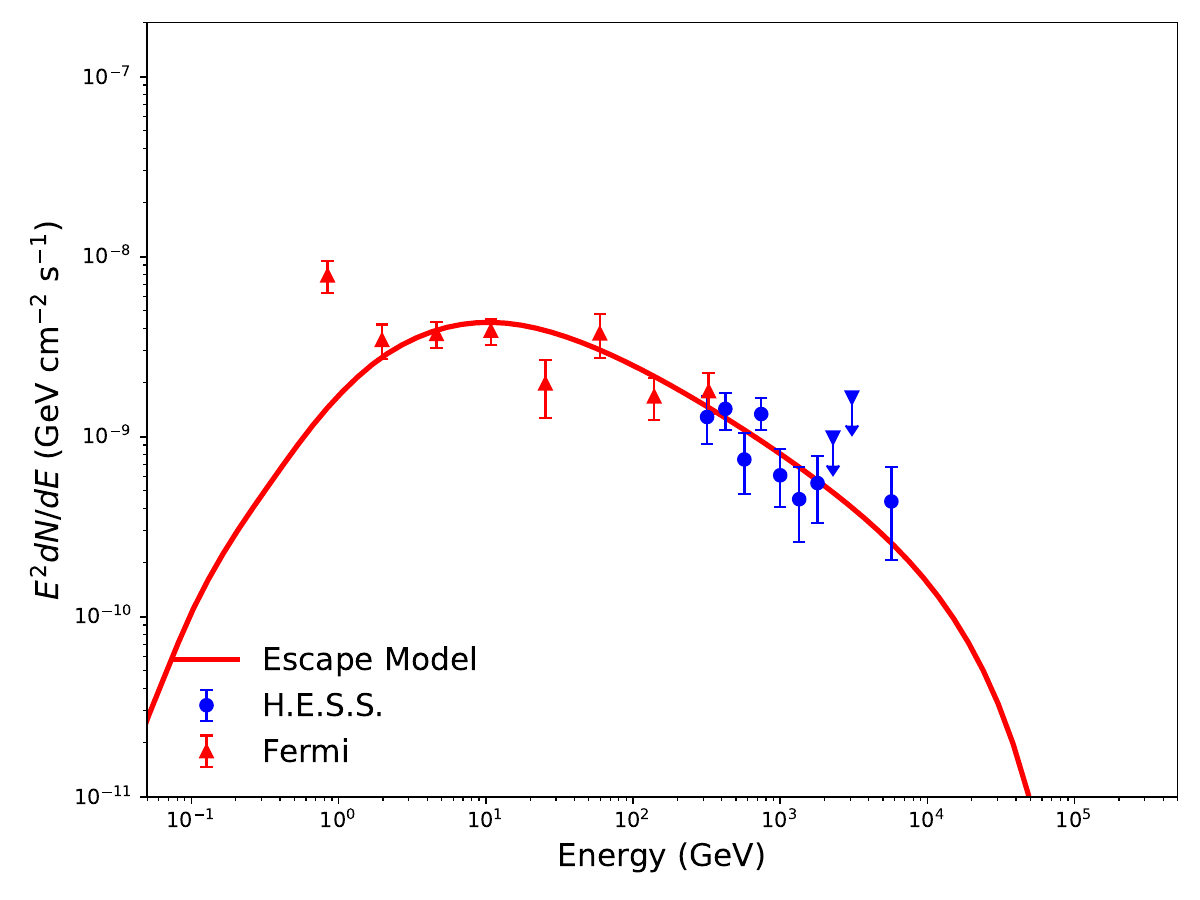}
     \includegraphics[width=1.\linewidth]{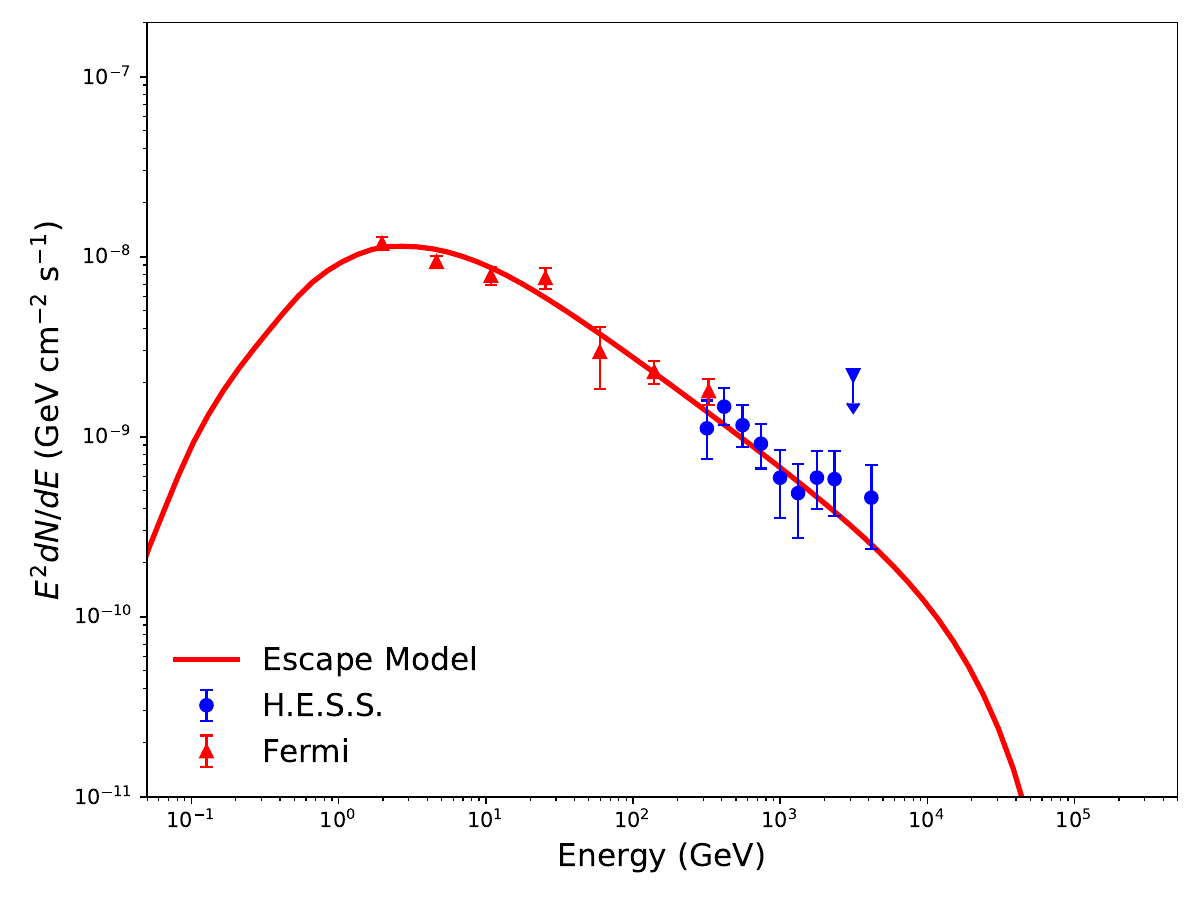}
    \caption{Similar to Fig.~\ref{fig:w51}, we compare the modeled gamma-ray emission from CRs within the SNR shell with the Fermi-LAT and H.E.S.S. data. The middle panel shows the fit to the gamma-ray emission from region A, and the lower panel shows the fit for region B, both based on the escape model.}
    \label{fig:w28}
\end{figure}

For W28, we adopt a shell radius of \(13\ {\rm pc}\) in our modeling. 
Our results are summarized in Fig.~\ref{fig:w28}. 
For W28, our modeling reveals that particle acceleration at the shock has ceased, and the observed gamma-ray emission is dominated by escaped particles interacting with a molecular cloud of mass \(5\times10^4\,M_\odot\) \cite{2013MNRAS.429.1643N}. Adopting an injected maximum energy of \(172\) TeV with a temporal decay of \(t^{-2.97}\), we find that the current acceleration capability becomes negligible, consistent with this picture. The inferred CR acceleration efficiency is \(\eta_{\rm cr} \sim 3\%\). These results indicate that W28 has entered a phase where essentially all accelerated particles have already escaped.

For the gamma-ray emission from regions A and B, we analyzed $>$0.1~GeV Fermi-LAT data during the time period from 2008-08-04 15:43:36 (UTC) to 2026-03-04 05:36:35 (UTC). We first used all the sources within $20^{\circ}$ listed in the Fermi 16-year source catalog \citep{fl16y} to construct the source model. There are four catalog sources within regions A and B. We found that replacing these four catalog sources with the extended templates of A and B improved the $\Delta\mathrm{AIC}$ by 12.8. When analyzing data above 5~GeV, the $\Delta\mathrm{AIC}$ improved by 70.8. At energies above 5~GeV, the point-spread function (PSF) is smaller, and contamination from surrounding sources is reduced. Therefore, we adopt the updated source model containing the extended templates of regions A and B to extract the gamma-ray spectra of these two extended sources.
To explain the emission from the outer regions, region A requires a molecular cloud of mass \(6\times10^4\,M_\odot\) located at a distance of \(46\) pc from the remnant, while region B requires a cloud of mass \(4\times10^4\,M_\odot\) at \(31\) pc. 

In our model, the gamma-ray emission from W28 is dominated entirely by escaped CRs interacting with molecular clouds, consistent with the H.E.S.S. finding that W28 has entered a radiative phase in which most of its CRs have escaped into the surrounding interstellar medium \cite{2008A&A...481..401A}.

\section{\label{sec:nu} High-energy neutrinos}
\begin{figure}[h!]
    \centering
    \includegraphics[width=\linewidth]{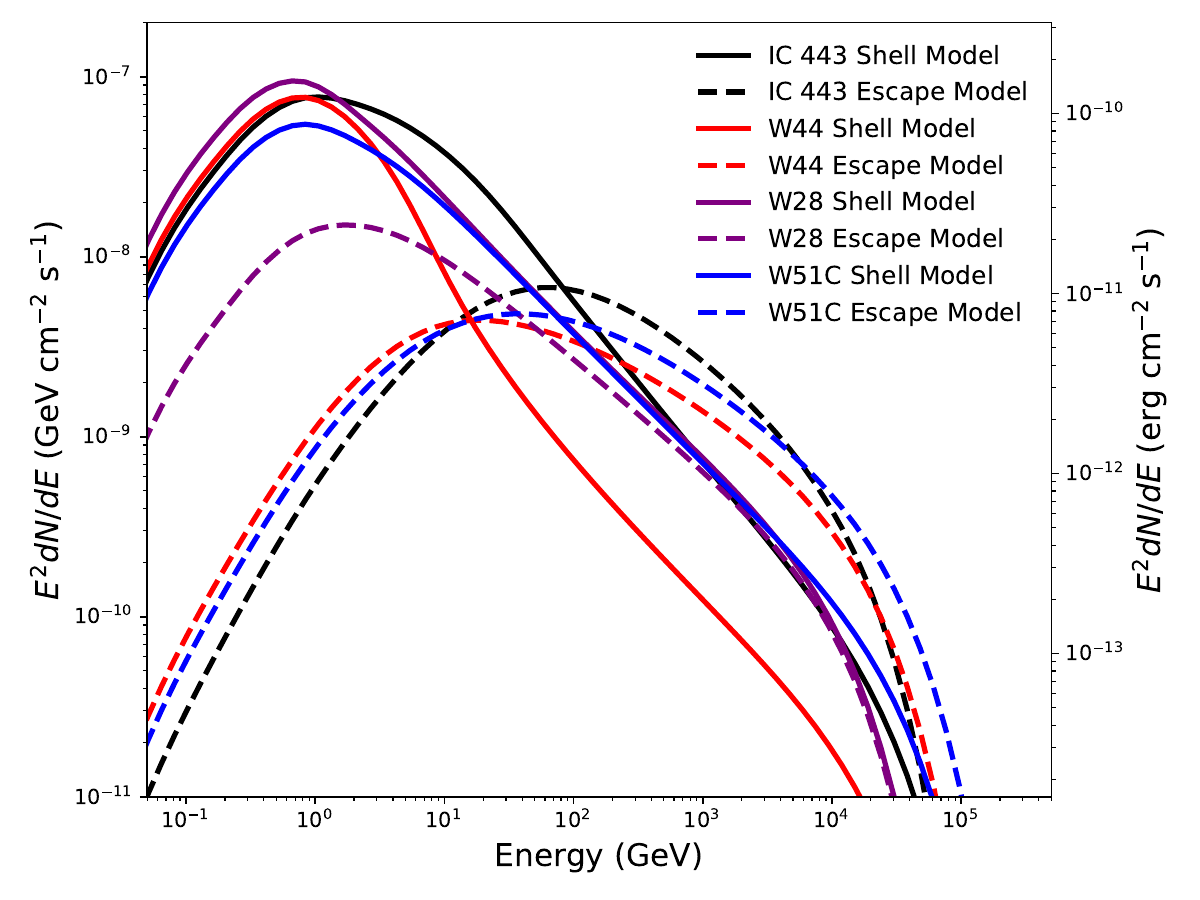}
    \caption{Modeled neutrino energy spectrum from CRs within the SNR shell (solid lines) and escaped-CRs interacting with surrounding molecular clouds (dashed lines) for the four SNRs considered in this work: W51C, IC443, W44, and W28.} 
    \label{fig:neutrino}
\end{figure}

In Fig.~\ref{fig:neutrino}, we show the corresponding all-flavor neutrino fluxes from these SNRs, including contributions from both the shell and escape models.
Our results demonstrate that accounting for escaped CRs can substantially increase the neutrino output from SNRs. 
We find that the neutrino flux above TeV energies is significantly enhanced when the contributions from previously escaped CRs are taken into account. 
In particular, for W44, the neutrino flux from escaped CRs is more than an order of magnitude higher than that from the shell component. 
For IC~443 and W51C, the fluxes above TeV are also enhanced by a factor of a few, while for W28 the escaped CR flux is comparable to that of the shell predictions.
This distinction is important when comparing with previous estimates \citep{2026ApJ...999...86S}, which convert gamma-ray flux to neutrino flux directly following the relation $\phi_\nu \propto \phi_\gamma$ and consider only CRs inside SNR, omitting the escaped component. 
Consequently, sources with faint shell emission, such as W44, classified as Tier~2 in those works (likely hadronic but lacking observed TeV emission), may in fact be promising targets for neutrino telescopes once the neutrinos from escaped-CRs are properly accounted for.

\begin{table*}[!ht]
\caption{\label{tab:params}
Physical parameters used in the MCMC analysis. Fixed parameters are listed together with the fitted parameters, for which the posterior mean values and corresponding $1\sigma$ credible intervals are shown.
}
\begin{ruledtabular}
\begin{tabular}{lccccccccc}
SNR & $\mathcal{E}_{\rm SN}$ [$10^{51}\rm~erg$] & $\tau_{\rm SN}$ [kyr] & $d$ [kpc] & $\alpha$ & $\delta$ & $\chi$ & $P_M$~($\mathrm{TeV}\,c^{-1}$) & $\eta_{\rm cr}$\footnote{$M_{\rm cl} = 2.1\times 10^4 M_\odot$ (W51C),  $10^3 M_\odot$ (IC 443), $ 10^3 M_\odot$ (G189.6+3.3), $5\times10^3 M_\odot$ (W44), $5\times 10^4 M_\odot$ (W28)} \\
\hline
W51C  & 3.6 & 30 & 5.5 &$4.26^{+0.09}_{-0.08}$ & $3.50^{+0.31}_{-0.38}$ & $0.11^{+0.05}_{-0.04}$ & $164^{+345}_{-110}$ & 0.11 &  \\
W51C~\footnote{The gamma-ray emission measured by Fermi-LAT
and LHAASO originates from a common component.}   & 3.6 & 30 & 5.5& $4.00^{+0.005}_{-0.002}$ &$3.89^{+0.08}_{-0.10}$& $0.29^{+0.03}_{-0.03}$& $543^{+70}_{-59}$ & 0.04
\\
IC 443 & 1.0 & 10& 1.5 & $4.34^{+0.03}_{-0.04}$  & $3.01^{+0.26}_{-0.29}$  & $0.04^{+0.01}_{-0.01}$  & $375^{+327}_{-183}$  & 0.04   \\
G189.6+3.3 & 1.0 & 100& 1.5 & $4.03^{+0.04}_{-0.02}$  & $2.49^{+1.01}_{-1}$  & $0.0002^{+0.0008}_{-0.0001}$  & $56^{+27}_{-13}$  & 0.02 \\
W44   & 5.0 & 20  & 2.2 & $4.25^{+0.06}_{-0.08}$& $1.82^{+0.34}_{-0.36}$ & $0.18^{+0.05}_{-0.04}$ & $105^{+385}_{-83}$& 0.01  \\
W28   &1.0 &35 & 1.9 & $4.21^{+0.05}_{-0.05}$ & $2.97^{+0.48}_{-0.46}$&$0.13^{+0.07}_{-0.04}$ & $172^{+393}_{-128}$& 0.03 
\end{tabular}
\end{ruledtabular}
\end{table*}

\section{\label{sec:dis} Discussion and Summary}
The fitted parameters of the four SNRs are summarized in Table~\ref{tab:params}. We find that the fitted spectral indices of the accelerated CRs are limited to a narrow range of $\alpha \sim 4.2-4.3$, which is steeper than the prediction of the test-particle DSA scenario for strong shocks.
Such spectral steepening could be expected in a refined DSA theory that accounts for the back-reaction of CRs~\citep{2012JCAP...07..038C}.
The maximum proton energies are constrained to $E_{p,\rm max}\sim 100$--$300~\rm TeV$ (see Table~\ref{tab:params}), although substantial uncertainties allow values ranging from $\sim 50$ TeV to $\sim 500$ TeV. This energy range is consistent with theoretical expectations from nonlinear CR-induced instabilities, which can amplify the magnetic field and thereby facilitate acceleration to higher energies~\cite{1983A&A...125..249L, 2004MNRAS.353..550B}. It is also consistent with the maximum CR energies adopted for the first population in the two-component framework of Ref.~\cite{Aharonian:2026tzf}.
The diffusion coefficient governing CR escape from SNRs is found to be lower than the standard Galactic value, with typical values about an order of magnitude below the Galactic average, suggesting suppressed diffusion in the vicinity of SNRs~\cite{Nava:2016szf}.

Our results for W51C reveal two possible interpretations of the gamma-ray data. In the first scenario, the emission observed by Fermi-LAT and LHAASO is attributed to a single component, namely CRs within the SNR shell interacting with shock-swept molecular clouds. While this shell-only scenario can reproduce the overall spectral shape, it requires a hard injection spectrum with \(\alpha \sim 4\). We therefore favor the second scenario, in which the gamma-rays detected by Fermi-LAT and MAGIC originate from CRs inside the SNR shell interacting with the shocked cloud material, while the LHAASO-detected UHE emission arises from CRs that have escaped from the SNR at earlier epochs and subsequently interact with a distant molecular cloud.
For IC 443, the C0 component is well reproduced by the shell-only model. For the extended C1 component, a shell-only interpretation is also possible if attributed to an older neighboring SNR G189.6+3.3. However, this scenario requires a hard spectral index of \(\alpha \sim 4\), which is again somewhat extreme. We therefore consider an alternative interpretation in which C1 originates from escaped CRs interacting with a molecular cloud of mass \(6.8\times10^{3}\,M_{\odot}\) at a distance of \(18\) pc, although this model provides a poorer fit at low energies. Given the difficulty in achieving a fully satisfactory fit with hadronic models, a leptonic contribution to C1 cannot be excluded~\cite{2026PhRvL.136p1002C}.

The escaped-CR scenario investigated in this work shares a similar physical picture with the model proposed by Ref.~\cite{Ohira_2010}.
In Ref.~\cite{Ohira_2010}, the broken power-law spectrum observed in gamma-rays arises from different escape behaviors of high and low energy CRs in the case that the SNR is embedded within a molecular cloud. 
Higher energy CRs escape gradually over time, yielding a steepened spectrum, while lower energy CRs cannot escape before the shock reaches the cloud boundary.
When the shock interacts with the dense cloud, the scattering waves that confine CRs are damped by the neutral gas, causing all low-energy CRs to escape simultaneously, retaining the original acceleration spectrum. The two spectral segments join at a break momentum determined by the shock-cloud interaction. 
Our modeling approach is different.
First, we treat the key physical parameters as free variables and constrain them through MCMC fitting to the multi-wavelength data. We simultaneously fit the spectral index, the maximum momentum, the diffusion coefficient normalization, and the acceleration efficiency. 
Note Ref.~\cite{Ohira_2010} assumes that the SNRs are PeV CR accelerators, with a maximum energy extending above a few PeV. 
Second, we adopt a scenario where the molecular cloud serves as a passive external target that does not affect the SNR evolution or the CR escape process. All CRs escape from the SNR according to their own time-dependent escape history, without the collective escape mechanism proposed by \cite{Ohira_2010}. Third, the broken power-law spectrum in our model arises from the superposition of two physically distinct components: CRs within the shell and escaped CRs, rather than a single population modified by different escape behaviors. 
Despite these differences, both studies support that the diffusion coefficient near the SNR is significantly suppressed compared to the Galactic average, indicating that the presence of the SNR strongly modifies the surrounding interstellar medium and enhances CR confinement in the vicinity of the remnant.

\medskip
B.T.Z. is supported in China by National Key R\&D program of China under the grant 2024YFA1611402. 
This work was also supported by the National Natural Science Foundation of China (Grants No. 12475114).
During the final stage of this work, we became aware of an independent study by Ref.~\cite{Sunny:2026jnz}, which also concluded that the UHE emission from W51 complex observed by LHAASO is likely produced by escaped CRs interacting with ambient gas.

\bibliographystyle{apsrev4-1}
\bibliography{refer.bib}

\end{document}